\documentclass[conference]{IEEEtran}
\IEEEoverridecommandlockouts
\usepackage{cite}
\usepackage{amsmath,amssymb,amsfonts}
\usepackage{algorithmic}
\usepackage{graphicx}
\usepackage{textcomp}
\usepackage{xcolor}
\def\BibTeX{{\rm B\kern-.05em{\sc i\kern-.025em b}\kern-.08em
		T\kern-.1667em\lower.7ex\hbox{E}\kern-.125emX}}
\usepackage[binary-units]{siunitx}
\usepackage[toc,section=chapter,nonumberlist,acronym]{glossaries}
\usepackage{tikz,pgfplots}
\usepackage{svg}
\usepackage{pgf-umlsd}
\usepackage[hyphens]{url}
\usepackage{breakurl}
\usepackage[hyperindex,breaklinks]{hyperref}
\usepackage{array}
\usepackage{booktabs}
\usepackage{url}
\usepackage{balance}

\usetikzlibrary{positioning,fit,arrows.meta,backgrounds,shapes,quotes,decorations.markings,calc,fit}
\usetikzlibrary{chains,automata}

\pgfplotsset{compat=1.16}

\newacronym{smartnic}{SmartNIC}{smart network interface card}
\newacronym{sdn}{SDN}{software-defined networking}
\newacronym{mat}{MAT}{match-action table}
\newacronym{ran}{RAN}{radio access network}
\newacronym{rss}{RSS}{receive-side scaling}
\newacronym{hdl}{HDL}{hardware description language}
\newacronym{rtl}{RTL}{register-transfer level}
\newacronym{tos}{TOS}{type of service}
\newacronym{mac}{MAC}{media access control}
\newacronym{phy}{PHY}{physical layer}
\newacronym{dma}{DMA}{direct memory access}
\newacronym{pcie}{PCIe}{peripheral component interconnect express}
\newacronym{vct}{VCT}{virtual cut-through}

\newcommand\copyrighttext{%
	\footnotesize \copyright 2024 IEEE. Personal use of this material is permitted. Permission from IEEE must be obtained for all other uses, in any current or future media, including reprinting/republishing this material for advertising or promotional purposes, creating new collective works, for resale or redistribution to servers or lists, or reuse of any copyrighted component of this work in other works.
}
\newcommand\copyrightnotice{%
	\begin{tikzpicture}[remember picture,overlay]
	\node[anchor=south,yshift=10pt] at (current page.south) {\fbox{\parbox{\dimexpr\textwidth-\fboxsep-\fboxrule\relax}{\copyrighttext}}};
	\end{tikzpicture}%
}

\begin{document}
	
	\title{%
		FlexCross: High-Speed and Flexible Packet Processing via a Crosspoint-Queued Crossbar%
		\thanks{We acknowledge the financial support from the Bavarian Ministry of Economic Affairs, Regional Development and Energy in the context of the project "6G Future Lab Bavaria".}%
		\copyrightnotice
	}
	
	\author{\IEEEauthorblockN{Klajd Zyla, Marco Liess, Thomas Wild, Andreas Herkersdorf}
\IEEEauthorblockA{\textit{Technical University of Munich, Germany}\\
\{klajd.zyla, marco.liess, thomas.wild, herkersdorf\}@tum.de}}
	
	\maketitle
	
	\begin{abstract}
    The fast pace at which new online services emerge leads to a rapid surge in the volume of network traffic. A recent approach that the research community has proposed to tackle this issue is in-network computing, which means that network devices perform more computations than before. As a result, processing demands become more varied, creating the need for flexible packet-processing architectures. State-of-the-art approaches provide a high degree of flexibility at the expense of performance for complex applications, or they ensure high performance but only for specific use cases. In order to address these limitations, we propose FlexCross. This flexible packet-processing design can process network traffic with diverse processing requirements at over 100 Gbit/s on FPGAs. Our design contains a crosspoint-queued crossbar that enables the execution of complex applications by forwarding incoming packets to the required processing engines in the specified sequence. The crossbar consists of distributed logic blocks that route incoming packets to the specified targets and resolve contentions for shared resources, as well as memory blocks for packet buffering. We implemented a prototype of FlexCross in Verilog and evaluated it via cycle-accurate register-transfer level simulations. We also conducted test runs with real-world network traffic on an FPGA. The evaluation results demonstrate that FlexCross outperforms state-of-the-art flexible packet-processing designs for different traffic loads and scenarios. The synthesis results show that our prototype consumes roughly 21\% of the resources on a Virtex XCU55 UltraScale+ FPGA.
\end{abstract}

\begin{IEEEkeywords}
    Network hardware, In-network computing, FPGA, Interconnect, Crossbar
\end{IEEEkeywords}

	\section{Introduction}

As our interconnected global landscape expands, an ever-growing set of online services is becoming prevalent, including video conferencing, extended reality, remote control, and various mobile applications \cite{src}. Consequently, there is a notable surge in network traffic, which imposes heavier processing demands on data centers and presents challenges in meeting application-specific bandwidth requirements and round-trip delay bounds \cite{MLE}. The primary bottlenecks that constrain CPU performance are power consumption and memory-access latency \cite{10.1145/3126908.3126970}. A recent technology that addresses this challenge is in-network computing \cite{9499874}. In-network computing signifies a paradigm shift where computational tasks are executed within the network infrastructure, diverging from the exclusive reliance on conventional server-based computing models. Within this framework, network devices like switches, routers, and \glspl{smartnic} are outfitted with specialized hardware and software to perform complex computational tasks, enhancing their traditional data-forwarding function. By delegating computational tasks to network devices, in-network computing helps to minimize data movement, reduce latency, and increase the overall system performance.

Due to the higher number of tasks network devices can execute, processing requirements are becoming more diverse, generating a demand for flexible packet-processing architectures. Some methods simplify the implementation of network applications by providing frameworks that allow developers to program at higher abstraction levels, e.g., \glspl{mat} \cite{10.1145/2872362.2872367}, extended finite state machines \cite{225996}, and Linux's eXpress Data Path \cite{258973}. Although these approaches provide higher flexibility than traditional NICs, they are not well-suited for complex computations. Other methods implement traditional protocol stacks in hardware, e.g., transport protocols \cite{arashloo2020enabling}, TCP/IP stack \cite{7577319}, and RDMA \cite{10.1145/3342195.3387519}. \Glspl{smartnic} that use such methods deploy hardware accelerators to reduce the processing time to a minimum for resource-intensive tasks, e.g., encryption, decompression, I/O virtualization, connection tracking, and congestion control \cite{bluefield}. Typically, these accelerators are integrated into a pipeline according to the sequence in which their functions must be executed. While this method proves efficient when all packet streams (flows) follow the same sequence of pipeline stages, it lacks inherent support for flows with distinct processing requirements. An example would be a \gls{smartnic} that receives both encrypted and unencrypted packets, compressed and uncompressed packets, and packets that need to be decrypted and decompressed in a different order. Running such applications on CPU cores would reduce the throughput and sacrifice the speedup achieved by accelerators.

In order to tackle these challenges, we present FlexCross. This flexible packet-processing design can process packets with different processing requirements at over \SI{100}{\giga\bit\per\second} on FPGAs. FlexCross attaches several processing engines to a shared crossbar-based interconnect. The crossbar can forward packets to the required processing engines in the specified sequence, thus enabling the execution of various complex applications. We instantiate multiple parallel units inside the corresponding processing engine for processing units that do not have enough capacity to process incoming packets at the line rate. The crossbar contains distributed logic blocks that route incoming packets to the specified destinations and resolve resource contentions, as well as memory blocks placed at the crosspoints for packet buffering. The execution sequence can be reconfigured in runtime depending on the requirements of each flow. Since FlexCross is modular and implemented on an FPGA, the number and types of processing engines can be modified as required. This property allows our design to deploy emerging protocols and applications.

We implemented a prototype of FlexCross in synthesizable \gls{hdl} code and evaluated it via cycle-accurate \gls{rtl} simulations. We also tested our design with real-world traffic on a NetFPGA-SUME \cite{netfpga}. The evaluation results demonstrate that FlexCross can process packets of arbitrary size with different processing requirements at the line rate. Furthermore, it outperforms state-of-the-art crossbar-based packet-processing designs regarding throughput by a factor of up to two. We also synthesized and implemented our prototype on an Alveo U55C High Performance Compute Card \cite{alveo-u55c}. The interconnect of our design scales quadratically in terms of chip area, but it still uses fewer queues than state-of-the-art crossbar switches.

In section \ref{rel_work}, we point out the main features of state-of-the-art crossbar switches and describe two flexible packet-processing designs that we use as a baseline for comparison in the evaluation. In section \ref{design}, we describe the main components of our design and point out how it addresses the limitations of existing approaches. In section \ref{eval}, we describe the simulation setup used to evaluate our design, discuss the evaluation results, and show the resource usage. Finally, we conclude the paper and briefly examine future work in section \ref{concl}.

	\section{Related Work}
\label{rel_work}

In this section, we summarize the state of the art regarding crossbar switches and point out the benefits and drawbacks of the most common approaches. Furthermore, we describe two flexible packet-processing designs that we use as a baseline for comparison in the evaluation.

\subsection{Crossbar Switches}
Since crossbars provide fast, simultaneous data transfer among different initiator-target pairs, they are the primary choice in high-bandwidth switches and routers. However, as the number of ports and the traffic rate increases, developing switches that achieve the required throughput becomes challenging. Early approaches in the literature are typically classified into three categories: output-queued (OQ), input-queued (IQ), and combined input- and output-queued (CIOQ) switches. Most of the approaches that fall into these categories employ centralized scheduling. The main factors that distinguish them apart from the location of the queues are the speedup with which the crossbar operates compared with the bandwidth of the input/output ports and the scheduling algorithm, which decides when and where each incoming packet must be forwarded.

Due to the lack of buffer space on the input side, OQ $N$x$N$ switches must operate with a speedup of $N$ to cope with a situation when packets coming from $N$ input ports must be forwarded to the same output port. Hence, OQ switches do not scale well and are thus considered impractical. IQ $N$x$N$ switches with a single per-port FIFO queue do not have to operate with a speedup of $N$, but they either suffer from head-of-line (HOL) blocking or need schedulers with a high time complexity to achieve \SI{100}{\percent} throughput. HOL blocking is the phenomenon that occurs when packets whose output port is free are blocked by packets whose output port is busy. In order to avoid HOL blocking, virtual output-queued (VOQ) switches are used, which, however, need significantly larger buffer space. CIOQ switches can achieve nearly \SI{100}{\percent} throughput while operating with a speedup of as low as 2 \cite{chuang1999matching}, but they still employ complex centralized scheduling. Minkenberg and Engbersen \cite{minkenberg2000combined} apply distributed two-stage scheduling with low complexity and deploy shared memory in the CIOQ switch. However, the bottlenecks in their switch architecture are the shared-memory interconnection complexity and the output queue bandwidth. In summary, all the approaches above suffer in one or more aspects: high speedup, HOL blocking, complex scheduling, and large queue size. Furthermore, they employ fixed-size packet scheduling, potentially wasting bandwidth and requiring packet segmentation and reassembly.

The recent advances in VLSI technology have enabled the integration of on-chip memory into the crossbar switch, which in turn has led to the emergence of buffered crossbar switches of two main types, combined input-crosspoint-queued (CICQ) switches \cite{pan2008localized} and combined input-crosspoint-output-queued (CICOQ) switches \cite{katevenis2004variable}. Output queues are usually necessary when the switch operates with a speedup. At the cost of larger buffer space, such switches employ distributed and variable-size packet scheduling, thus addressing the drawbacks of CIOQ switches. Katevenis et al. \cite{katevenis2004variable} propose a CICQ switch architecture that achieves nearly \SI{100}{\percent} throughput without speedup. However, they deploy per-port schedulers both on the input and the output side.

\subsection{Flexible Packet Processing}
\label{subsec:flex}
PANIC \cite{panic} represents a high-performance, adaptable NIC designed to accommodate flows with diverse processing demands. Employing a CIOQ crossbar \cite{becker2012efficient}, it dynamically routes incoming packets to the available processing units in a reconfigurable order. PANIC deploys multiple parallel units to achieve the required throughput for low-bandwidth computing units. A \gls{mat}-based parser analyzes incoming packets and determines to which computing units the crossbar must forward them and in which sequence based on the corresponding flows. This information is encoded in packet descriptors and forwarded ahead of the packet data sequentially. A centralized scheduler that contains hardware-based priority queues called PIFOs \cite{10.1145/2934872.2934899} privileges high-priority packets in high-load conditions. PANIC applies push/pull scheduling and load-aware steering. Once a computing unit has finished processing a packet, it directly forwards it to the next required unit instead of sending it back to the scheduler. This method reduces the per-packet latency and the bandwidth demands on the interconnect. However, if the ingress queue of the computing unit that receives the packet is full, the receiving unit returns the packet to the scheduler. In this case, the scheduler provides load-aware steering by forwarding incoming packets to the computing unit with the smallest number of packets in its ingress queue. If all eligible computing units are occupied, the scheduler buffers the packets until a unit becomes available.

FlexPipe \cite{zyla2023flexpipe} represents a versatile packet-processing pipeline tailored for high-performance \glspl{smartnic}. Similarly to PANIC, it utilizes a \gls{mat}-based parser to derive the necessary processing sequence for each packet based on its associated flow. However, FlexPipe distinguishes itself by transmitting the generated metadata alongside the packet data via a separate channel, resulting in enhanced throughput. By organizing various offloads into a pipeline and supplementing them with adaptable forwarding logic, FlexPipe enables offload bypassing depending on the processing needs of incoming packets. This point-to-point interconnect requires less chip area compared to a crossbar. Each offload module features a load balancer responsible for directing incoming packets to the offload unit with the fewest data in its ingress queue, thus minimizing queuing time. This method differs from PANIC, which redirects traffic to another offload unit only when the ingress queue of the current unit is full. FlexPipe deploys as many offload units as necessary to process packets at the line rate for each offload type. At the egress of each offload, a traffic arbiter governs packet forwarding, prioritizing packets stored in the queue with the highest fill level to balance the available space. Additionally, FlexPipe enables packet recirculation through the offload pipeline via two traffic controllers, facilitating not only offload bypassing but also the modification of activation sequences. However, recirculated packets consume on-chip bandwidth and require a trade-off between throughput and flexibility.

	\section{Proposed Design}
\label{design}

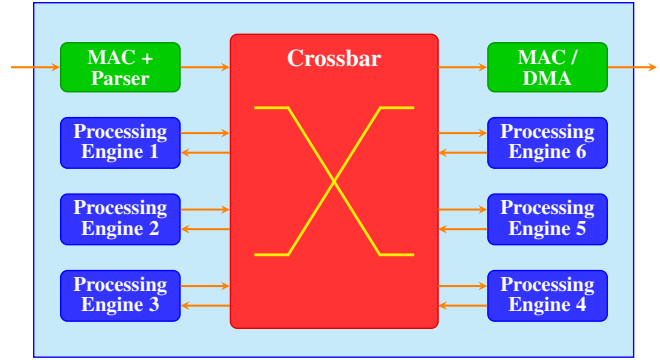
\begin{figure}[tb]
	\centering
	\resizebox{\linewidth}{!}{
		\begin{tikzpicture}
	\node[rectangle, draw, blue, fill=cyan!20, thick, text width=12cm, text height=7cm] at (4.35,-2.3) () {};
	\node[rectangle, draw, rounded corners, green!60!black, fill=green!80!black, text=white, thick, text centered, text width= 2.2cm, minimum height= 1cm, font=\bf] (preprocess) {\large{MAC + Parser}};
	\draw[-stealth, orange, very thick] ([xshift=-1cm]preprocess.west)--node[midway,above]{}(preprocess.west);
	\draw[-stealth, orange, very thick] (preprocess.east)--node[midway,above]{}([xshift=1cm]preprocess.east);
	\node[rectangle, draw, rounded corners, blue, fill=blue!80, text=white, thick, text centered, below=0.5cm of preprocess, text width= 2.2cm, minimum height= 1cm, font=\bf] (serv1) {\large{Processing Engine 1}};
	\draw[-stealth, orange, very thick] ([yshift=2mm]serv1.east)--node[midway,above]{}([xshift=1cm,yshift=2mm]serv1.east);
	\draw[-stealth, orange, very thick] ([xshift=1cm,yshift=-2mm]serv1.east)--node[midway,above]{}([yshift=-2mm]serv1.east);
	\node[rectangle, draw, rounded corners, blue, fill=blue!80, text=white, thick, text centered, below=0.5cm of serv1, text width= 2.2cm, minimum height= 1cm, font=\bf] (serv2) {\large{Processing Engine 2}};
	\draw[-stealth, orange, very thick] ([yshift=2mm]serv2.east)--node[midway,above]{}([xshift=1cm,yshift=2mm]serv2.east);
	\draw[-stealth, orange, very thick] ([xshift=1cm,yshift=-2mm]serv2.east)--node[midway,above]{}([yshift=-2mm]serv2.east);
	\node[rectangle, draw, rounded corners, blue, fill=blue!80, text=white, thick, text centered, below=0.5cm of serv2, text width= 2.2cm, minimum height= 1cm, font=\bf] (serv3) {\large{Processing Engine 3}};
	\draw[-stealth, orange, very thick] ([yshift=2mm]serv3.east)--node[midway,above]{}([xshift=1cm,yshift=2mm]serv3.east);
	\draw[-stealth, orange, very thick] ([xshift=1cm,yshift=-2mm]serv3.east)--node[midway,above]{}([yshift=-2mm]serv3.east);
	\node[rectangle, draw, rounded corners, red!90!black, fill=red!80, text=white, thick, text centered, below right= -1.2cm and 1cm of preprocess, text width= 4cm, minimum height= 6cm, font=\bf] (cross) {};
	\node[white, thick, above= -1cm of cross, minimum width= 2cm, minimum height= 1cm, font=\bf] (crosstext) {\Large{Crossbar}};
	\draw[yellow, ultra thick] ([xshift=0.5cm,yshift=1.5cm]cross.west)--node[midway,above]{}([xshift=1.2cm,yshift=1.5cm]cross.west);
	\draw[yellow, ultra thick] ([xshift=0.5cm,yshift=-1.5cm]cross.west)--node[midway,above]{}([xshift=1.2cm,yshift=-1.5cm]cross.west);
	\draw[yellow, ultra thick] ([xshift=-1.2cm,yshift=1.5cm]cross.east)--node[midway,above]{}([xshift=-0.5cm,yshift=1.5cm]cross.east);
	\draw[yellow, ultra thick] ([xshift=-1.2cm,yshift=-1.5cm]cross.east)--node[midway,above]{}([xshift=-0.5cm,yshift=-1.5cm]cross.east);
	\draw[yellow, ultra thick] ([xshift=1.18cm,yshift=1.52cm]cross.west)--node[midway,above]{}([xshift=-1.19cm,yshift=-1.52cm]cross.east);
	\draw[yellow, ultra thick] ([xshift=1.18cm,yshift=-1.52cm]cross.west)--node[midway,above]{}([xshift=-1.18cm,yshift=1.52cm]cross.east);
	
	\node[rectangle, draw, rounded corners, green!60!black, fill=green!80!black, text=white, thick, text centered, right=6.25cm of preprocess, text width= 2.2cm, minimum height= 1cm, font=\bf] (dma) {\large{MAC / DMA}};
	\draw[-stealth, orange, very thick] ([xshift=-1cm]dma.west)--node[midway,above]{}(dma.west);
	\draw[-stealth, orange, very thick] (dma.east)--node[midway,above]{}([xshift=1cm]dma.east);
	\node[rectangle, draw, rounded corners, blue, fill=blue!80, text=white, thick, text centered, right=6.25cm of serv1, text width= 2.2cm, minimum height= 1cm, font=\bf] (serv6) {\large{Processing Engine 6}};
	\draw[-stealth, orange, very thick] ([xshift=-1cm,yshift=2mm]serv6.west)--node[midway,above]{}([yshift=2mm]serv6.west);
	\draw[-stealth, orange, very thick] ([yshift=-2mm]serv6.west)--node[midway,above]{}([xshift=-1cm,yshift=-2mm]serv6.west);
	\node[rectangle, draw, rounded corners, blue, fill=blue!80, text=white, thick, text centered, right=6.25cm of serv2, text width= 2.2cm, minimum height= 1cm, font=\bf] (serv5) {\large{Processing Engine 5}};
	\draw[-stealth, orange, very thick] ([xshift=-1cm,yshift=2mm]serv5.west)--node[midway,above]{}([yshift=2mm]serv5.west);
	\draw[-stealth, orange, very thick] ([yshift=-2mm]serv5.west)--node[midway,above]{}([xshift=-1cm,yshift=-2mm]serv5.west);
	\node[rectangle, draw, rounded corners, blue, fill=blue!80, text=white, thick, text centered, right=6.25cm of serv3, text width= 2.2cm, minimum height= 1cm, font=\bf] (serv4) {\large{Processing Engine 4}};
	\draw[-stealth, orange, very thick] ([xshift=-1cm,yshift=2mm]serv4.west)--node[midway,above]{}([yshift=2mm]serv4.west);
	\draw[-stealth, orange, very thick] ([yshift=-2mm]serv4.west)--node[midway,above]{}([xshift=-1cm,yshift=-2mm]serv4.west);
\end{tikzpicture}
	}
	\caption{\label{fig:flexcross_arch}Block diagram of the architecture of FlexCross}
\end{figure}

In order to address the drawbacks of state-of-the-art crossbar switches and flexible NIC architectures, we propose FlexCross, the architecture of which is depicted in Fig. \ref{fig:flexcross_arch}. FlexCross runs at \SI{200}{\mega\hertz} and has a data width of \SI{512}{\bit}, thus achieving a bandwidth of \SI{102.4}{\giga\bit\per\second}. All modules exchange data via the AXI4-Stream protocol \cite{axis}.

\subsection{Overall Architecture}
The \textit{\gls{mac}} receives IP packets encapsulated in Ethernet frames from the \gls{phy}, performs layer 2 functions, and sends them to the \textit{Parser}. The Parser inspects incoming packets and generates metadata based on the extracted information. The metadata are transmitted along with the data stream as user-defined sideband information via the TUSER signal of the AXI4-Stream protocol, thus not affecting the bandwidth of the data path. They convey useful information about the packet to the subsequent design modules, such as the packet size, flow type, priority class, required task sequence, next required task, and timestamp. The flow type is derived from a specified set of fields in the header, such as the source/destination IP address or TCP/UDP port, and is mapped to a task sequence. This mapping can be reconfigured in runtime based on the emerging processing demands of each flow. The priority class is extracted from the \gls{tos} / priority field of the IPv4/IPv6 header. The Parser forwards each packet to the \textit{Crossbar}, deployed as a shared on-chip interconnect. The Crossbar steers incoming packets to the required \textit{Processing Engine}s in the specified sequence, thus enabling task chaining. This feature makes FlexCross modular and enables it to support various applications. We deploy six Processing Engines in our prototype, but their number and types can be modified as needed. In the last step, the packets are forwarded either to the \gls{mac} (in a switch/router), which sends them to the \gls{phy}, or to the \textit{\gls{dma}} Engine (in a NIC), which sends incoming packets to the attached server via the \gls{pcie}. A 7x7 Crossbar is sufficient to connect all the design modules.

\subsection{Processing Engines}
The Processing Engines perform the tasks required from the flows, e.g., checksum verification, en-/decryption, IPv4 routing, or network address translation (NAT). Suppose a single processing unit needs more bandwidth to process packets at the line rate. In that case, we instantiate multiple parallel units, which share the compute load of the incoming traffic. Moreover, we deploy a load balancer at the Processing Engine's ingress that monitors each processing unit's load and forwards packets to the least loaded unit \cite{zyla2023flexpipe}. A round-robin traffic arbiter at the egress resolves contentions when multiple processing units have finished processing their packets. Before packets leave the Processing Engine, the metadata field "next required task" is updated to the following task in the specified sequence.

We implement four types of processing units that perform simple packet header inspection/modification: firewall, NAT, IPv4 router, and load balancer. The firewall determines whether an incoming packet must be forwarded or dropped based on the source port in its TCP/UDP header. The NAT modifies the destination IP address of arriving packets based on a translation table. The IPv4 router selects the Ethernet port to which a packet must be forwarded based on the destination address in its IP header. The load balancer distributes the traffic among the available Ethernet ports by applying the round-robin policy. We also use two publicly available types of processing units. In total, we integrate six different processing engines running at \SI{200}{\mega\hertz} into FlexCross and attach them to the crossbar: checksum verification (CRC) \cite{crc}, firewall, NAT, en-/decryption (AES) \cite{aes}, IPv4 router and load balancer. The CRC engine has an input width of \SI{256}{\bit}, thus needing two processing units to process packets at the line rate. The AES engine has an input width of \SI{128}{\bit}, thus needing four processing units to process packets at the line rate.

\subsection{Crossbar}
\begin{figure}[tb]
	\centering
	\resizebox{\linewidth}{!}{
		\input{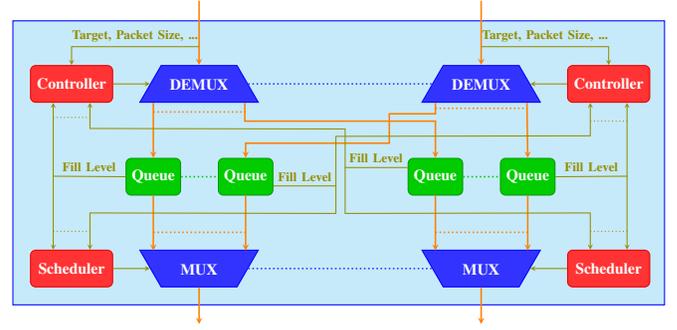}
	}
	\caption{\label{fig:crossbar_arch}Block diagram of the architecture of the Crossbar}
\end{figure}

We develop a crosspoint-queued crossbar to connect the Processing Engines. The location of the queues allows us to decouple the crossbar's input side from the output side and apply local scheduling for each output port. Fig. \ref{fig:crossbar_arch} represents the architecture of the Crossbar, which consists of data path blocks (in blue), control path blocks (in red), and memory blocks (in green). The data signals are in orange, while the control signals are in olive. Each Demultiplexer (DEMUX) receives incoming packets from the output port of the Parser / Processing Engine to which it is connected and forwards them to the corresponding FIFO queue \cite{verilog-axis} of the Multiplexer (MUX) that is connected to the input port of the target (another Processing Engine / MAC / DMA). We set the queue size to \SI{8}{\kilo\byte}. At least five \SI{1500}{\byte} packets, the maximum transmission unit for Ethernet, fit in a queue. We employ \gls{vct} switching, which means that we apply non-blocking packet-level flow control and read packet flits from each queue as soon as the corresponding MUX output is free, instead of waiting for the whole packet to arrive, as in store-and-forward switching. Each DEMUX has its own Controller, which generates control signals based on the target (next required task), the packet size, and the fill level of the corresponding queue. The fill level is tracked to determine whether the remaining queue space is enough to store the packet. Suppose the remaining space is smaller than the packet size (including the Ethernet header). In that case, the packet is dropped to avoid backpressure, which can lead to congestion propagation, non-work conservation, and deadlocks at high traffic loads \cite{gregoire2014capacity, zaidi2016back}. This feature is realized by asserting the TVALID signal transmitted to the target FIFO queue LOW and driving the TREADY signal to the input port HIGH. The evaluation shows that the chosen queue size is sufficient to keep the packet drop rate low, even in high-load scenarios. Each MUX fetches packets from the queues attached to its inputs and forwards them to the input port of the Processing Engine / MAC / DMA to which it is connected. The local Scheduler resolves contentions by determining the queue from which the next packet must be read based on the selected algorithm and the fill level of each queue. It is capable of making scheduling decisions every clock cycle. Hence, it allows back-to-back packets of minimum size to be forwarded without delay.

\subsection{Scheduler}
We implement three scheduling algorithms for our prototype: round robin (RR), longest queue first (LQF), and first come, first served (FCFS).

We implement the RR scheduler as a 7-input multiplexer, which polls the FIFO queues connected to the inputs of the corresponding MUX and selects the first one (queue $i$) that is not empty (TVALID is HIGH). After the MUX forwards the whole packet at the head of queue $i$ to the output, the scheduler polls the queues again by starting with queue $i+1$ or queue $0$ if $i=7$. This algorithm is simple and starvation-free.

We implement the LQF scheduler as a tree comparator. In the first round, the queues connected to the inputs of the corresponding MUX are compared with each other regarding their fill levels, and the queue with the highest fill level is selected in each pair. These queues are compared in the second round in the same clock cycle, and the queue with the highest fill level is selected in each newly formed pair again. This process continues until only one queue remains, which is then granted the right to forward a packet to the output of the corresponding MUX. This algorithm achieves a very low packet drop rate by selecting the queue with the highest fill level in every scheduling cycle. However, it is not starvation-free.

The FCFS scheduler consists of a FIFO queue and enqueuing and dequeuing logic. When a packet is stored in any of the queues connected to the inputs of the corresponding MUX, the enqueuing logic stores the index of the target queue to the closest empty cell to the head of the FIFO queue. When a scheduling cycle occurs, the dequeuing logic fetches the index at the head of the queue and propagates it as a control signal to the corresponding MUX. This algorithm is more complex than the ones above. However, since it ensures that packets leave the queues according to their arrival order, it reduces variations in queuing delay.

\subsection{Comparison to the State of the Art}
Our crossbar addresses the aspects in which OQ, IQ, and CIOQ crossbar switches suffer because it operates without speedup, prevents HOL blocking by deploying separate queues at the crosspoints for all possible input-output pairs, simplifies scheduling by distributing its logic, and uses fewer queues. In contrast to the most prominent crosspoint-queued switch \cite{katevenis2004variable}, we place no queues on the input side, deploy per-port schedulers only on the output side, and employ \gls{vct} switching. Hence, the crossbar deployed in FlexCross uses fewer hardware resources and scales better than state-of-the-art switches. Our crossbar uses $N^2$ queues, while \cite{minkenberg2000combined} uses $N^2+N$ queues, \cite{pan2008localized} uses $2*N^2+N$ queues, and \cite{katevenis2004variable} uses $2*N^2$ queues in total. FlexCross outperforms PANIC because the CIOQ crossbar deployed in PANIC has no virtual output queues and operates without speedup, leading to HOL blocking. Compared with FlexPipe, our design further reduces the per-packet latency by avoiding additional queuing time when bypassing multiple Processing Engines and achieves higher throughput in scenarios in which FlexPipe recirculates a large traffic share. However, it does not scale as well as FlexPipe since the number of wires and queues deployed in the crossbar is $N^2$. The claimed performance improvements are supported by the evaluation results.

	\section{Evaluation}
\label{eval}

We implemented a prototype of FlexCross in Verilog and compared it with PANIC \cite{panic}, FlexPipe \cite{zyla2023flexpipe}, and a CIOQ crossbar-based design \cite{becker2012efficient} in terms of throughput and latency for different traffic rates and scenarios. We obtained the measurements from cycle-accurate \gls{rtl} simulations in Vivado \cite{vivado}. We also ran tests with real-world traffic on a NetFPGA-SUME \cite{netfpga}. Furthermore, we synthesized and implemented FlexCross on an Alveo U55C High Performance Compute Card \cite{alveo-u55c}, which features a Virtex XCU55 UltraScale+ FPGA \cite{virtex-ultra}.

\subsection{Simulation Setup}
\label{subsec:sim_setup}
In order to evaluate the architectures above, we developed a traffic generator and a traffic sink in Verilog, which interface with each design via the AXI4-Stream protocol. The traffic generator injects packets of variable size, allocates each of them a flow type and a priority class, and configures the mapping of flow types to task sequences in the parser. It probabilistically determines when a packet must be sent based on the specified traffic rate, which leads to short-term load variations, thus making the traffic pattern more realistic. Furthermore, it generates flow types and packet sizes according to a uniform distribution. The traffic sink receives processed packets from the design and measures the throughput, mean, minimum, and maximum per-packet latency. The throughput is the rate at which the design forwards data to the traffic sink in \SI{}{\giga\bit\per\second}, while the per-packet latency is the time in \SI{}{\micro\second} that passes from the moment the packet enters the parser until the moment it arrives at the traffic sink.

We simulated an open-source prototype of PANIC \cite{panic_osdi20_artifact}, which we modified to address two scalability issues. Since the crossbar deployed in PANIC \cite{becker2012efficient} does not support more than eight initiators/targets, we integrate all offload units of the same type into a single module, as in FlexCross. Another limitation is that the scheduler supports up to two priority queues, but six are needed (one for each offload). In order to circumvent this issue, we avoid additional trips to the centralized scheduler by determining the offload unit to which a packet must be forwarded at the ingress of the offload. Hence, we simulated an optimized prototype of PANIC. We also simulated a prototype of FlexPipe \cite{zyla2023flexpipe} and a design based on the CIOQ crossbar used in PANIC \cite{becker2012efficient}. We integrate the CIOQ crossbar into the same overall architecture as FlexCross to verify the exclusive impact of the interconnect on the performance. All simulated designs have the same bandwidth (\SI{102.4}{\giga\bit\per\second}) and deploy the same number of processing units for each processing engine. We use as many units as required to achieve \SI{100}{\percent} throughput, i.e., one unit for firewall, NAT, IPv4 router, and load balancer, two units for CRC, and four units for AES.

In each simulation run, we injected 52,000 synthetic IP packets of variable size and equal priority into each design, each associated with a flow type. We start the measurement after the traffic sink has received 1,000 packets and stop it after it has received 51,000 packets to ensure that the interconnect load is steady throughout the measurement phase. The mean latency did not vary significantly during each simulation run. When a packet must be dispatched, the traffic generator randomly selects a frame size from the set \{\SI{64}{\byte}, \SI{128}{\byte}, \SI{256}{\byte}, \SI{512}{\byte}, \SI{1024}{\byte}, \SI{1518}{\byte}\} according to a uniform distribution.

In the first scenario, the traffic generator randomly assigns one out of four available flow types to the packet according to a uniform distribution. We map each flow type to a different task sequence:

\begin{itemize}
	\item Flow 1: CRC $\rightarrow$ firewall $\rightarrow$ AES $\rightarrow$ load balancer $\rightarrow$ NAT
	\item Flow 2: firewall $\rightarrow$ NAT $\rightarrow$ AES $\rightarrow$ IPv4 router
	\item Flow 3: CRC $\rightarrow$ AES $\rightarrow$ IPv4 router
	\item Flow 4: CRC $\rightarrow$ load balancer
\end{itemize}

In the second scenario, we randomize the task sequence assigned to each injected packet. Each task sequence includes all processing engines in a randomly determined order. This scenario might not be realistic for the deployed processing engines. For example, CRC might always need to be performed first, or the IPv4 router and the load balancer might never be required to be applied to the same flow. However, it allows us to test our design in conditions where the crossbar reaches the highest possible load, leading to frequent contentions for processing engines and high queuing delays.

\subsection{Simulation Results}
\label{subsec:sim_res}
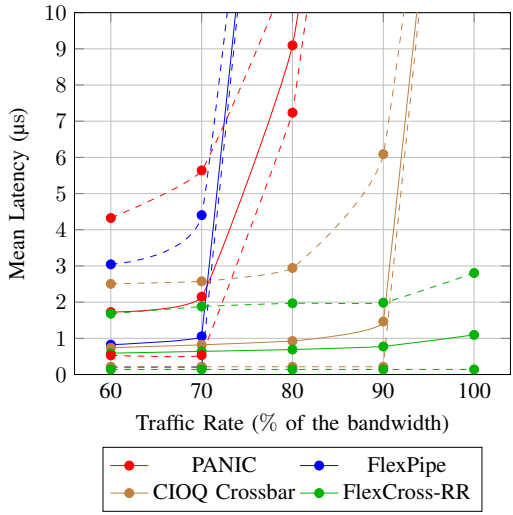
\begin{figure}[tbp]
	\centering
	\resizebox{.8\linewidth}{!}{
		\begin{tikzpicture}
	\begin{axis}[
		xlabel={Traffic Rate (\SI{}{\percent} of the bandwidth)},
		symbolic x coords={60, 70, 80, 90, 100},
		ylabel={Mean Latency (\SI{}{\micro\second})},
		ymin=0, ymax=10,
		xtick=data,
		ytick={0,1,2,3,4,5,6,7,8,9,10},
		legend columns=2,
		legend style={at={(0.5,-0.2)},anchor=north},
		xmajorgrids=true,
		ymajorgrids=true,
		]
		
		\addplot[
		color=red,
		mark=*,
		mark options={solid},
		smooth,
		tension=0.2
		]
		coordinates {
			(60,1.725)(70,2.150)(80,9.095)(90,25)(100,25)
		};
		\addlegendentry{PANIC}
		
		\addplot[
		color=blue,
		mark=*,
		mark options={solid},
		smooth,
		tension=0.02
		]
		coordinates {
			(60,0.825)(70,1.055)(80,25)(90,25)(100,25)
		};
		\addlegendentry{FlexPipe}
		
		\addplot[
		color=brown,
		mark=*,
		mark options={solid},
		smooth,
		tension=0.05
		]
		coordinates {
			(60,0.740)(70,0.820)(80,0.925)(90,1.465)(100,25)
		};
		\addlegendentry{CIOQ Crossbar}
		
		\addplot[
		color=green!70!black,
		mark=*,
		mark options={solid},
		smooth,
		tension=0.2
		]
		coordinates {
			(60,0.590)(70,0.640)(80,0.690)(90,0.775)(100,1.095)
		};
		\addlegendentry{FlexCross-RR}
		
		\addplot[
		color=red,
		style=dashed,
		mark=*,
		mark options={solid},
		smooth,
		tension=0.05
		]
		coordinates {
			(60,0.530)(70,0.535)(80,7.235)(90,25)(100,25)
		};
		
		\addplot[
		color=red,
		style=dashed,
		mark=*,
		mark options={solid},
		smooth,
		tension=0.2
		]
		coordinates {
			(60,4.325)(70,5.640)(80,11.550)(90,25)(100,25)
		};
		
		\addplot[
		color=blue,
		style=dashed,
		mark=*,
		mark options={solid},
		smooth,
		tension=0
		]
		coordinates {
			(60,0.205)(70,0.205)(80,25)(90,25)(100,25)
		};
		
		\addplot[
		color=blue,
		style=dashed,
		mark=*,
		mark options={solid},
		smooth,
		tension=0.2
		]
		coordinates {
			(60,3.045)(70,4.405)(80,25)(90,25)(100,25)
		};
		
		\addplot[
		color=brown,
		style=dashed,
		mark=*,
		mark options={solid},
		smooth,
		tension=0
		]
		coordinates {
			(60,0.215)(70,0.215)(80,0.215)(90,0.215)(100,25)
		};
		
		\addplot[
		color=brown,
		style=dashed,
		mark=*,
		mark options={solid},
		smooth,
		tension=0.2
		]
		coordinates {
			(60,2.505)(70,2.575)(80,2.945)(90,6.090)(100,25)
		};
		
		\addplot[
		color=green!70!black,
		style=dashed,
		mark=*,
		mark options={solid},
		smooth,
		tension=0.2
		]
		coordinates {
			(60,0.140)(70,0.140)(80,0.140)(90,0.140)(100,0.140)
		};
		
		\addplot[
		color=green!70!black,
		style=dashed,
		mark=*,
		mark options={solid},
		smooth,
		tension=0.2
		]
		coordinates {
			(60,1.680)(70,1.880)(80,1.970)(90,1.985)(100,2.805)
		};
		
	\end{axis}
\end{tikzpicture}
	}
	\caption{\label{fig:latency}Mean per-packet latency when receiving traffic associated with four flow types at different rates measured in PANIC, the CIOQ crossbar-based design, FlexPipe, and FlexCross with RR scheduling. The dashed lines show the minimum/maximum measured latency.}
\end{figure}

Fig. \ref{fig:latency} shows the mean, minimum, and maximum latency measured in PANIC, FlexPipe, the CIOQ crossbar-based design, and FlexCross with RR scheduling in the first scenario when receiving packets at different rates. FlexCross can process traffic at any rate and thus provide line-rate packet processing at \SI{100}{\giga\bit\per\second}. As expected, the mean per-packet latency is the lowest for a traffic rate of \SI{60}{\percent} (\SI{0.59}{\micro\second}) and the highest for a rate of \SI{100}{\percent} (\SI{1.095}{\micro\second}). The latency increases as the traffic rate increases due to the higher frequency of contentions for processing engines, which leads to longer queuing time. FlexCross drops approximately \SI{0.2}{\percent} of the injected packets only when receiving traffic at \SI{100}{\percent} of the bandwidth. On the other hand, FlexPipe, PANIC, and the CIOQ crossbar-based design generate backpressure and saturate at rates of \SI{70}{\percent}, \SI{80}{\percent} and \SI{90}{\percent}, respectively. Moreover, FlexCross achieves a lower mean, minimum, and maximum per-packet latency than the other designs for any traffic rate. FlexPipe delivers lower performance than FlexCross because it needs to recirculate packets belonging to flow 1. The main reason for the lower performance of PANIC and the CIOQ crossbar-based design compared with FlexCross is that the deployed CIOQ crossbar has no virtual output queues and operates without speedup, which leads to HOL blocking.

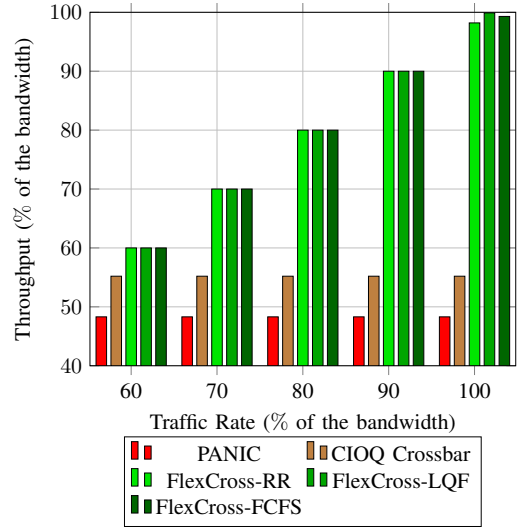
\begin{figure}[tbp]
	\centering
	\resizebox{.8\linewidth}{!}{
		\begin{tikzpicture}
	\begin{axis}[
		xlabel={Traffic Rate (\SI{}{\percent} of the bandwidth)},
		symbolic x coords={60, 70, 80, 90, 100},
		ylabel={Throughput (\SI{}{\percent} of the bandwidth)},
		enlarge x limits=0.12,
		ymin=40, ymax=100,
		ybar,
		xtick=data,
		ytick={40,50,60,70,80,90,100},
		legend columns=2,
		legend style={at={(0.5,-0.2)},anchor=north},
		xmajorgrids=true,
		ymajorgrids=true,
		bar width=1.7mm
		]
		
		\addplot[
		fill=red,
		]
		coordinates {
			(60,48.3)(70,48.3)(80,48.3)(90,48.3)(100,48.3)
		};
		\addlegendentry{PANIC}
		
		\addplot[
		fill=brown,
		]
		coordinates {
			(60,55.2)(70,55.2)(80,55.2)(90,55.2)(100,55.2)
		};
		\addlegendentry{CIOQ Crossbar}
		
		\addplot[
		fill=green!90!black,
		]
		coordinates {
			(60,60)(70,70)(80,80)(90,90)(100,98.2)
		};
		\addlegendentry{FlexCross-RR}
		
		\addplot[
		fill=green!65!black,
		]
		coordinates {
			(60,60)(70,70)(80,80)(90,90)(100,99.9)
		};
		\addlegendentry{FlexCross-LQF}
		
		\addplot[
		fill=green!40!black,
		]
		coordinates {
			(60,60)(70,70)(80,80)(90,90)(100,99.3)
		};
		\addlegendentry{FlexCross-FCFS}
		
	\end{axis}
\end{tikzpicture}
	}
	\caption{\label{fig:throughput}Throughput in \% of the bandwidth when receiving traffic at different rates mapped to randomly generated task sequences achieved by PANIC, the CIOQ crossbar-based design, and FlexCross with three different schedulers}
\end{figure}

Fig. \ref{fig:throughput} shows the throughput achieved by PANIC, the CIOQ crossbar-based design, and FlexCross with three different schedulers in the second scenario when receiving packets at different rates. FlexCross achieves the maximum throughput up to a traffic rate of at least \SI{90}{\percent} of the bandwidth without dropping any packets. For a traffic rate of \SI{100}{\percent} FlexCross-RR, -LQF and -FCFS attain a throughput / drop rate of \SI{98.2}{\percent}/\SI{1}{\percent}, \SI{99.9}{\percent}/\SI{0.1}{\percent} and \SI{99.3}{\percent}/\SI{0.3}{\percent}, respectively. The packet drop rate can be decreased by increasing the queue size. The results show that FlexCross can provide line-rate packet processing at \SI{100}{\giga\bit\per\second} even when the crossbar reaches the maximum load. By contrast, PANIC and the CIOQ crossbar-based design suffer the consequences of HOL blocking, achieving a maximum throughput of only \SI{48.3}{\percent} and \SI{55.2}{\percent}, respectively, while FlexPipe does not support scenarios in which flows vary significantly in the required task sequence due to the high packet recirculation rate.

\begin{figure}[tbp]
	\centering
	\resizebox{.925\linewidth}{!}{
		\begin{tikzpicture}
	\begin{axis}[
		xlabel={Traffic Rate (\SI{}{\percent} of the bandwidth)},
		symbolic x coords={60, 70, 80, 90, 100},
		ylabel={Mean Latency (\SI{}{\micro\second})},
		ymin=0, ymax=22,
		xtick=data,
		ytick={0,2,4,6,8,10,12,14,16,18,20},
		legend columns=3,
		legend style={at={(0.5,-0.2)},anchor=north},
		xmajorgrids=true,
		ymajorgrids=true,
		]
		
		\addplot[
		color=red,
		mark=*,
		mark options={solid},
		smooth,
		tension=0.2
		]
		coordinates {
			(60,1.110)(70,1.320)(80,1.625)(90,2.360)(100,4.965)
		};
		\addlegendentry{FlexCross-RR}
		
		\addplot[
		color=blue,
		mark=*,
		mark options={solid},
		smooth,
		tension=0.2
		]
		coordinates {
			(60,1.225)(70,1.510)(80,1.925)(90,2.810)(100,9.430)
		};
		\addlegendentry{FlexCross-LQF}
		
		\addplot[
		color=green!70!black,
		mark=*,
		mark options={solid},
		smooth,
		tension=0.2
		]
		coordinates {
			(60,1.110)(70,1.315)(80,1.615)(90,2.310)(100,6.615)
		};
		\addlegendentry{FlexCross-FCFS}
		
		\addplot[
		color=red,
		style=dashed,
		mark=*,
		mark options={solid},
		smooth,
		tension=0.2
		]
		coordinates {
			(60,0.475)(70,0.480)(80,0.495)(90,0.560)(100,1.405)
		};
		
		\addplot[
		color=red,
		style=dashed,
		mark=*,
		mark options={solid},
		smooth,
		tension=0.2
		]
		coordinates {
			(60,3.725)(70,4.330)(80,4.460)(90,6.615)(100,12.785)
		};
		
		\addplot[
		color=blue,
		style=dashed,
		mark=*,
		mark options={solid},
		smooth,
		tension=0.2
		]
		coordinates {
			(60,0.475)(70,0.475)(80,0.515)(90,0.675)(100,2.405)
		};
		
		\addplot[
		color=blue,
		style=dashed,
		mark=*,
		mark options={solid},
		smooth,
		tension=0.2
		]
		coordinates {
			(60,4.185)(70,5.095)(80,6.000)(90,10.025)(100,20.665)
		};
		
		\addplot[
		color=green!70!black,
		style=dashed,
		mark=*,
		mark options={solid},
		smooth,
		tension=0.2
		]
		coordinates {
			(60,0.475)(70,0.475)(80,0.500)(90,0.655)(100,3.600)
		};
		
		\addplot[
		color=green!70!black,
		style=dashed,
		mark=*,
		mark options={solid},
		smooth,
		tension=0.2
		]
		coordinates {
			(60,2.630)(70,2.975)(80,3.450)(90,5.110)(100,9.980)
		};
		
		\draw[orange, thick] (axis cs:60,0.688) -- (axis cs:100,0.688);
		
	\end{axis}
\end{tikzpicture}
	}
	\caption{\label{fig:latency_random}Mean per-packet latency when receiving traffic at different rates mapped to randomly generated task sequences measured in FlexCross with three different schedulers. The dashed lines show the minimum/maximum measured latency, while the orange indicates the mean processing delay.}
\end{figure}
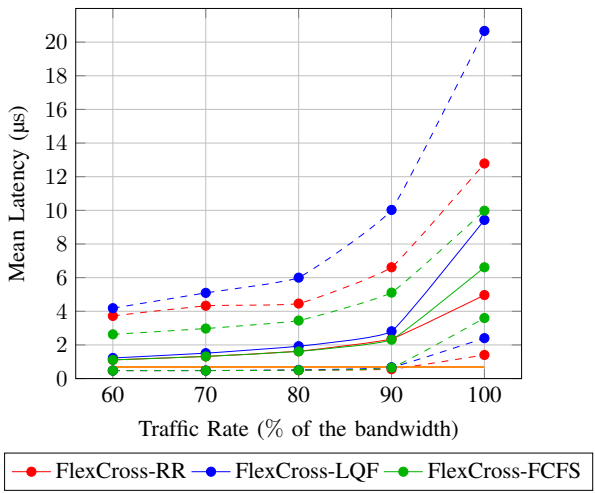

Fig. \ref{fig:latency_random} shows the mean, minimum, and maximum latency measured in FlexCross with three different schedulers in the second scenario when receiving packets at different rates. FlexCross-RR, -LQF, and -FCFS achieve a similar mean/minimum per-packet latency up to a traffic rate of \SI{90}{\percent} of the bandwidth. The discrepancy is most significant for a rate of \SI{100}{\percent} because the schedulers differ in their packet drop rate. LQF has the highest per-packet latency but also the lowest drop rate. FCFS achieves the lowest maximum latency for any traffic rate, which makes this scheduler a better alternative for flows that require tight latency bounds. RR is the least complex scheduler regarding time and area, but it still achieves high throughput and low latency. It has a higher drop rate because it is not tailored to minimize the variation of queue fill levels. The orange line at the bottom of the plot indicates the mean processing delay over the selected packet sizes, including the crossbar traversal latency (four clock cycles). The mean queuing delay is the difference from the mean latency. Since the frequency of resource contentions increases with the traffic rate, the queuing delay also goes up.

\subsection{Test Results}
In order to validate the simulation results and show the capability of our design to achieve very low per-packet latency at the line rate when processing real-world network traffic, we implemented FlexCross with RR scheduling on an FPGA and injected IP packets via an Ethernet port. We used the open-source network tester FlueNT10G \cite{oeldemann2018fluent10g} to replay ten sets of traces of \SI{60}{\second} duration each (more than 600 million packets) from the CAIDA UCSD Anonymized Internet Traces 2019 Dataset \cite{passive_2019_pcap}. Moreover, we extract performance measurements from the traffic sink via the Integrated Logic Analyzer (ILA) IP core from Xilinx \cite{ila}. The used dataset contains traces collected on a \SI{10}{\giga\bit\per\second} link (mean traffic rate of \SI{45}{\percent}) from the Equinix NYC high-speed monitor. We replayed five sets of traces corresponding to direction A (Sao Paulo to New York), which have a mean packet size of \SI{880}{\byte}, and five sets of traces corresponding to direction B (New York to Sao Paulo), which have a mean packet size of \SI{380}{\byte}. We increased the traffic rate by augmenting the real-world traces with synthetic traffic, which is achieved by inserting packets of variable size in the gaps between the captured packets (without modifying the original timestamps). We implemented FlexCross on a NetFPGA-SUME \cite{netfpga}, which contains four Ethernet ports operating at \SI{10}{\giga\bit\per\second}. For a fair evaluation, we proportionally reduced the bandwidth of our design by scaling the clock frequency down to \SI{20}{\mega\hertz}. As in the simulation, we tested FlexCross with two task sequence allocations. In the first scenario, we assign packets that contain specific TCP/UDP destination ports to one of the four flow types mentioned in subsection \ref{subsec:sim_setup} using the modulo operator, while in the second scenario, we allocate a randomly generated task sequence that includes all processing engines to each incoming packet. Table \ref{tab:test_results} shows the mean, minimum, and maximum per-packet latency measured during the simulation with synthetic traffic and the test runs in both scenarios. The amount of dropped packets was less than \SI{0.1}{\percent}, and the mean throughput was equal to the traffic rate.

\begin{table}[t]
	\centering
	\caption{\label{tab:test_results}Mean, minimum, and maximum latency in clock cycles measured during the simulation with synthetic traffic and the test runs on an FPGA in two scenarios}
	\begin{tabular}{*8c}
		\toprule
		Evaluation & Load & \multicolumn{3}{c}{Fixed Task Sequence} & \multicolumn{3}{c}{Random Task Sequence} \\
		\midrule
		{} & {} & Mean & Min. & Max. & Mean & Min. & Max. \\
		Simulation & 70\% & 128 & 28 & 376 & 264 & 96 & 866 \\
		Test dirA & 70\% & 157 & 28 & 817 & 387 & 95 & 2747 \\
		Simulation & 95\% & 169 & 28 & 498 & 641 & 205 & 1599 \\
		Test dirB & 95\% & 152 & 28 & 921 & 652 & 96 & 2756 \\
		\bottomrule
	\end{tabular}
\end{table}

\subsection{Resource Usage}
\begin{figure}[tbp]
	\centering
	\resizebox{.85\linewidth}{!}{
		\begin{tikzpicture}
	\begin{axis}[
		xlabel={Resource type},
		symbolic x coords={LUT, LUTRAM, FF, BRAM},
		ylabel={Utilization (\SI{}{\percent})},
		enlarge x limits=0.17,
		ymin=0, ymax=100,
		ybar,
		xtick=data,
		ytick={0,10,20,30,40,50,60,70,80,90,100},
		legend columns=2,
		legend style={at={(0.5,-0.2)},anchor=north},
		xmajorgrids=true,
		ymajorgrids=true,
		bar width=2.3mm
		]
		
		\addplot
		coordinates {
			(LUT,10.65)(LUTRAM,2.40)(FF,8.41)(BRAM,32.79) 
		};
		\addlegendentry{Processing Units}
		
		\addplot
		coordinates {
			(LUT,12.28)(LUTRAM,2.61)(FF,9.23)(BRAM,41.39)
		};
		\addlegendentry{FlexPipe}
		
		\addplot
		coordinates {
			(LUT,2.33)(LUTRAM,0)(FF,0.96)(BRAM,21.43)
		};
		\addlegendentry{Crossbar 7x7}
		
		\addplot
		coordinates {
			(LUT,11.62)(LUTRAM,3.67)(FF,3.79)(BRAM,83.23)
		};
		\addlegendentry{Crossbar 14x14}
		
		\addplot[
		fill=green!70!black
		]
		coordinates {
			(LUT,14.05)(LUTRAM,2.61)(FF,9.93)(BRAM,56)
		};
		\addlegendentry{FlexCross 7x7}
		
	\end{axis}
\end{tikzpicture}
	}
	\caption{\label{fig:resource_usage}Resource usage of FlexPipe and FlexCross on an Alveo U55C High Performance Compute Card. LUT = lookup table, LUTRAM = lookup table RAM, FF = flip-flop, BRAM = block RAM}
\end{figure}
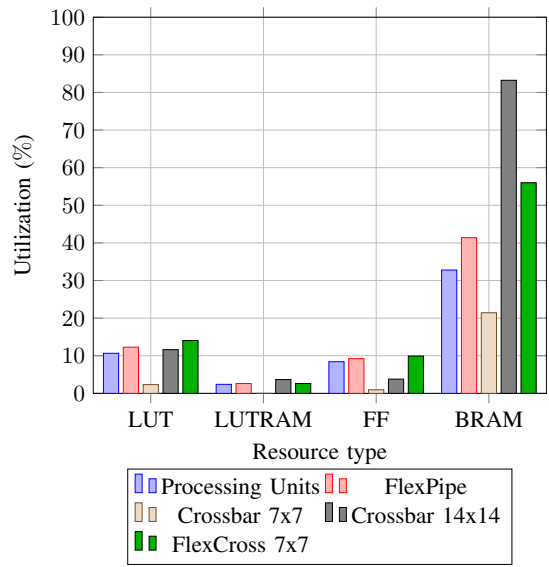

We integrated FlexPipe and FlexCross with RR scheduling into the AMD OpenNIC Shell \cite{opennic_shell}, which contains MAC and DMA subsystems, and ran Synthesis and Implementation in Vivado 2022.2, using the Alveo U55C High Performance Compute Card \cite{alveo-u55c} as target device. FlexCross passes timing for a bandwidth of \SI{102.4}{\giga\bit\per\second} (\SI{512}{\bit} x \SI{200}{\mega\hertz}). Fig. \ref{fig:resource_usage} depicts the share of the available FPGA resources FlexPipe and FlexCross consume. In order to quantify the overhead that the interconnect introduces, we break the total resource usage down into the amount occupied by the processing units and the crossbar (7x7). We also determined the resource usage of a 14x14 crossbar to show how the interconnect scales. A Virtex XCU55 UltraScale+ FPGA \cite{virtex-ultra} contains \SI{1303680}{} LUTs, \SI{36.7}{\mega\bit} of LUTRAM, \SI{2607360}{} FFs and \SI{70.9}{\mega\bit} of BRAM. FlexCross uses only \SI{1.77}{\percent} more LUTs, \SI{0.7}{\percent} more FFs and \SI{14.61}{\percent} more BRAM than FlexPipe. The crossbar (7x7) utilizes only \SI{16}{\percent} of the consumed LUTs, \SI{9}{\percent} of the consumed FFs and \SI{38}{\percent} of the consumed BRAM. As expected, when the number of ports doubles, the resource usage of the crossbar roughly quadruples, thus exhibiting quadratic scalability. The BRAM blocks utilize only a quarter of their total capacity (512 words) due to the chosen queue depth (128 words). Hence, we could quadruple the queue size without increasing BRAM utilization, thus potentially achieving an even higher throughput and lower packet drop rate.

	\section{Conclusion and Outlook}
\label{concl}

In order to tackle the challenges faced by state-of-the-art crossbar switches and flexible NIC architectures, we proposed FlexCross. This flexible packet-processing design can process packets with different processing requirements at over \SI{100}{\giga\bit\per\second} on FPGAs. FlexCross consists of a crosspoint-queued crossbar that connects several processing engines. Each processing engine has multiple application-specific processing units, a load balancer, and a round-robin traffic arbiter if a single unit cannot process back-to-back packets. We implemented a prototype of FlexCross with six processing engines in Verilog and evaluated it via cycle-accurate \gls{rtl} simulations. The simulation results demonstrate that FlexCross can process packets of arbitrary size with different processing requirements at the line rate. We also synthesized and implemented our design on an Alveo U55C High Performance Compute Card \cite{alveo-u55c} and showed its resource usage. Moreover, we performed test runs with real-world traffic on an FPGA. The synthesis results indicate that our prototype consumes about 21\% of the resources on a Virtex XCU55 UltraScale+ FPGA \cite{virtex-ultra}. In future work, we plan to consider the knowledge of the types of applications that run on switches and \glspl{smartnic}. By considering the sequence in which tasks can be executed, we can develop a partially connected crossbar, thus reducing the chip area. We also intend to enhance the per-port schedulers with priority awareness to fulfill the quality-of-service requirements of different traffic classes.

	\balance
	
	\bibliographystyle{IEEEtran}
	\bibliography{IEEEabrv,references}

\begin{thebibliography}{10}
\providecommand{\url}[1]{#1}
\csname url@samestyle\endcsname
\providecommand{\newblock}{\relax}
\providecommand{\bibinfo}[2]{#2}
\providecommand{\BIBentrySTDinterwordspacing}{\spaceskip=0pt\relax}
\providecommand{\BIBentryALTinterwordstretchfactor}{4}
\providecommand{\BIBentryALTinterwordspacing}{\spaceskip=\fontdimen2\font plus
\BIBentryALTinterwordstretchfactor\fontdimen3\font minus
  \fontdimen4\font\relax}
\providecommand{\BIBforeignlanguage}[2]{{%
\expandafter\ifx\csname l@#1\endcsname\relax
\typeout{** WARNING: IEEEtran.bst: No hyphenation pattern has been}%
\typeout{** loaded for the language `#1'. Using the pattern for}%
\typeout{** the default language instead.}%
\else
\language=\csname l@#1\endcsname
\fi
#2}}
\providecommand{\BIBdecl}{\relax}
\BIBdecl

\bibitem{src}
J.~Ang \emph{et~al.}, \emph{Decadal Plan for Semiconductors}.\hskip 1em plus
  0.5em minus 0.4em\relax Semiconductor Research Corporation, 2021.

\bibitem{MLE}
{Missing Link Electronics}, ``{Network Function Accelerators, FACs, NICs and
  SmartNICs},''
  \url{https://www.missinglinkelectronics.com/function-accelerators/}.

\bibitem{10.1145/3126908.3126970}
T.~Hoefler, S.~Di~Girolamo, K.~Taranov, R.~E. Grant, and R.~Brightwell,
  ``{sPIN}: High-performance streaming processing in the network,'' in
  \emph{Proceedings of the International Conference for High Performance
  Computing, Networking, Storage and Analysis}, ser. SC '17.\hskip 1em plus
  0.5em minus 0.4em\relax New York, NY, USA: Association for Computing
  Machinery, 2017.

\bibitem{9499874}
S.~Di~Girolamo, A.~Kurth, A.~Calotoiu, T.~Benz, T.~Schneider, J.~Beránek,
  L.~Benini, and T.~Hoefler, ``A {RISC-V} in-network accelerator for flexible
  high-performance low-power packet processing,'' in \emph{2021 ACM/IEEE 48th
  Annual International Symposium on Computer Architecture (ISCA)}, 2021, pp.
  958--971.

\bibitem{10.1145/2872362.2872367}
A.~Kaufmann, S.~Peter, N.~K. Sharma, T.~Anderson, and A.~Krishnamurthy, ``High
  performance packet processing with flexnic,'' in \emph{Proceedings of the
  Twenty-First International Conference on Architectural Support for
  Programming Languages and Operating Systems}, ser. ASPLOS '16.\hskip 1em plus
  0.5em minus 0.4em\relax New York, NY, USA: Association for Computing
  Machinery, 2016, p. 67–81.

\bibitem{225996}
S.~Pontarelli, R.~Bifulco, M.~Bonola, C.~Cascone, M.~Spaziani, V.~Bruschi,
  D.~Sanvito, G.~Siracusano, A.~Capone, M.~Honda, F.~Huici, and G.~Bianchi,
  ``{FlowBlaze}: Stateful packet processing in hardware,'' in \emph{16th USENIX
  Symposium on Networked Systems Design and Implementation (NSDI 19)}.\hskip
  1em plus 0.5em minus 0.4em\relax Boston, MA: USENIX Association, Feb. 2019,
  pp. 531--548.

\bibitem{258973}
M.~S. Brunella, G.~Belocchi, M.~Bonola, S.~Pontarelli, G.~Siracusano,
  G.~Bianchi, A.~Cammarano, A.~Palumbo, L.~Petrucci, and R.~Bifulco, ``{hXDP}:
  Efficient software packet processing on {FPGA} {NICs},'' in \emph{14th USENIX
  Symposium on Operating Systems Design and Implementation (OSDI 20)}.\hskip
  1em plus 0.5em minus 0.4em\relax USENIX Association, Nov. 2020, pp. 973--990.

\bibitem{arashloo2020enabling}
M.~T. Arashloo, A.~Lavrov, M.~Ghobadi, J.~Rexford, D.~Walker, and D.~Wentzlaff,
  ``Enabling programmable transport protocols in
  $\{$High-Speed$\}$$\{$NICs$\}$,'' in \emph{17th USENIX Symposium on Networked
  Systems Design and Implementation (NSDI 20)}, 2020, pp. 93--109.

\bibitem{7577319}
D.~Sidler, Z.~István, and G.~Alonso, ``Low-latency {TCP/IP} stack for data
  center applications,'' in \emph{2016 26th International Conference on Field
  Programmable Logic and Applications (FPL)}, 2016, pp. 1--4.

\bibitem{10.1145/3342195.3387519}
D.~Sidler, Z.~Wang, M.~Chiosa, A.~Kulkarni, and G.~Alonso, ``{StRoM}: Smart
  remote memory,'' in \emph{Proceedings of the Fifteenth European Conference on
  Computer Systems}, ser. EuroSys '20.\hskip 1em plus 0.5em minus 0.4em\relax
  New York, NY, USA: Association for Computing Machinery, 2020.

\bibitem{bluefield}
Nvidia, ``{NVIDIA BlueField-3 DPU},''
  \url{https://resources.nvidia.com/en-us-accelerated-networking-resource-library/datasheet-nvidia-bluefield?lx=LbHvpR&topic=networking-cloud}.

\bibitem{netfpga}
{AMD Xilinx}, ``{NetFPGA-SUME FPGA Development Board},''
  \url{https://www.xilinx.com/products/boards-and-kits/1-6ogkf5.html}.

\bibitem{alveo-u55c}
------, ``{Alveo U55C High Performance Compute Card},''
  \url{https://www.xilinx.com/products/boards-and-kits/alveo/u55c.html}.

\bibitem{chuang1999matching}
S.-T. Chuang, A.~Goel, N.~McKeown, and B.~Prabhakar, ``Matching output queueing
  with a combined input/output-queued switch,'' \emph{IEEE Journal on Selected
  Areas in Communications}, vol.~17, no.~6, pp. 1030--1039, 1999.

\bibitem{minkenberg2000combined}
C.~Minkenberg and T.~Engbersen, ``A combined input and output queued packet
  switched system based on prizma switch on a chip technology,'' \emph{IEEE
  Communications Magazine}, vol.~38, no.~12, pp. 70--77, 2000.

\bibitem{pan2008localized}
D.~Pan and Y.~Yang, ``Localized independent packet scheduling for buffered
  crossbar switches,'' \emph{IEEE Transactions on Computers}, vol.~58, no.~2,
  pp. 260--274, 2008.

\bibitem{katevenis2004variable}
M.~Katevenis, G.~Passas, D.~Simos, I.~Papaefstathiou, and N.~Chrysos,
  ``Variable packet size buffered crossbar (cicq) switches,'' in \emph{2004
  IEEE International Conference on Communications (IEEE Cat. No. 04CH37577)},
  vol.~2.\hskip 1em plus 0.5em minus 0.4em\relax IEEE, 2004, pp. 1090--1096.

\bibitem{panic}
J.~Lin, K.~Patel, B.~E. Stephens, A.~Sivaraman, and A.~Akella, ``{PANIC}: A
  {High-Performance} programmable {NIC} for multi-tenant networks,'' in
  \emph{14th USENIX Symposium on Operating Systems Design and Implementation
  (OSDI 20)}.\hskip 1em plus 0.5em minus 0.4em\relax USENIX Association, Nov.
  2020, pp. 243--259.

\bibitem{becker2012efficient}
D.~U. Becker, \emph{Efficient microarchitecture for network-on-chip
  routers}.\hskip 1em plus 0.5em minus 0.4em\relax Stanford University, 2012.

\bibitem{10.1145/2934872.2934899}
A.~Sivaraman, S.~Subramanian, M.~Alizadeh, S.~Chole, S.-T. Chuang, A.~Agrawal,
  H.~Balakrishnan, T.~Edsall, S.~Katti, and N.~McKeown, ``Programmable packet
  scheduling at line rate,'' in \emph{Proceedings of the 2016 ACM SIGCOMM
  Conference}, ser. SIGCOMM '16.\hskip 1em plus 0.5em minus 0.4em\relax New
  York, NY, USA: Association for Computing Machinery, 2016, p. 44–57.

\bibitem{zyla2023flexpipe}
K.~Zyla, M.~Liess, T.~Wild, and A.~Herkersdorf, ``{FlexPipe}: Fast, flexible
  and scalable packet processing for high-performance {SmartNICs},'' in
  \emph{2023 IFIP/IEEE 31st International Conference on Very Large Scale
  Integration (VLSI-SoC)}.\hskip 1em plus 0.5em minus 0.4em\relax IEEE, 2023,
  pp. 1--6.

\bibitem{axis}
{Arm Developer}, ``{AMBA 4 AXI4-Stream Protocol Specification},''
  \url{https://developer.arm.com/documentation/ihi0051/a}.

\bibitem{crc}
``{Ultimate CRC},'' \url{https://opencores.org/projects/ultimate_crc}.

\bibitem{aes}
``{AES},'' \url{https://opencores.org/projects/tiny_aes}.

\bibitem{verilog-axis}
A.~Forencich, ``{Verilog AXI Stream Components},''
  \url{https://github.com/alexforencich/verilog-axis}.

\bibitem{gregoire2014capacity}
J.~Gregoire, X.~Qian, E.~Frazzoli, A.~De~La~Fortelle, and T.~Wongpiromsarn,
  ``Capacity-aware backpressure traffic signal control,'' \emph{IEEE
  Transactions on Control of Network Systems}, vol.~2, no.~2, pp. 164--173,
  2014.

\bibitem{zaidi2016back}
A.~A. Zaidi, B.~Kulcs{\'a}r, and H.~Wymeersch, ``Back-pressure traffic signal
  control with fixed and adaptive routing for urban vehicular networks,''
  \emph{IEEE Transactions on Intelligent Transportation Systems}, vol.~17,
  no.~8, pp. 2134--2143, 2016.

\bibitem{vivado}
{AMD Xilinx}, ``{Vivado ML},''
  \url{https://www.xilinx.com/products/design-tools/vivado.html}.

\bibitem{virtex-ultra}
------, ``{Virtex UltraScale+},''
  \url{https://www.xilinx.com/products/silicon-devices/fpga/virtex-ultrascale-plus.html}.

\bibitem{panic_osdi20_artifact}
``{PANIC},''
  \url{https://bitbucket.org/uw-madison-networking-research/panic_osdi20_artifact/src/master/}.

\bibitem{oeldemann2018fluent10g}
A.~Oeldemann, T.~Wild, and A.~Herkersdorf, ``Fluent10g: A programmable
  fpga-based network tester for multi-10-gigabit ethernet,'' in \emph{2018 28th
  International Conference on Field Programmable Logic and Applications
  (FPL)}.\hskip 1em plus 0.5em minus 0.4em\relax IEEE, 2018, pp. 178--1787.

\bibitem{passive_2019_pcap}
``{The CAIDA UCSD Anonymized Internet Traces 2019 Dataset},''
  \url{https://catalog.caida.org/dataset/passive_2019_pcap}, dates used:
  2019-01-17. Accessed: 2024-02-07.

\bibitem{ila}
{AMD Xilinx}, ``{Integrated Logic Analyzer v6.2 LogiCORE IP Product Guide
  (PG172)},'' \url{https://docs.xilinx.com/v/u/en-US/pg172-ila}.

\bibitem{opennic_shell}
``{AMD OpenNIC Shell},'' \url{https://github.com/Xilinx/open-nic-shell}.

\end{thebibliography}
	
\end{document}